\documentclass[eqsecnum,aps,twocolumn,epsf,showpacs]{revtex4} % PH. REV.
\usepackage{graphicx}
\begin{document}

\title{Dissipation-managed  soliton  in
a quasi-one-dimensional Bose-Einstein condensate}

\author{Sadhan K. Adhikari\footnote{Electronic
address: adhikari@ift.unesp.br; \\
URL: http://www.ift.unesp.br/users/adhikari/}}
\affiliation
{Instituto de F\'{\i}sica Te\'orica, UNESP - S\~ao Paulo State University,
01.405-900 S\~ao Paulo, S\~ao Paulo, Brazil}
 
\date{\today}

\begin{abstract}
We use the time-dependent mean-field Gross-Pitaevskii
equation 
to study the formation of a dynamically-stabilized dissipation-managed 
bright soliton in a quasi-one-dimensional Bose-Einstein condensate (BEC). 
Because of three-body recombination of bosonic atoms to molecules, atoms 
are lost (dissipated) from a BEC. Such dissipation leads to the decay of a 
BEC soliton. We demonstrate by a perturbation procedure that an 
alimentation of atoms from an external source to the BEC may compensate 
for the dissipation loss and lead to a dynamically-stabilized soliton.  
The result of the analytical perturbation method is in excellent agreement 
with mean-field numerics.  It seems possible to obtain such a 
dynamically-stabilized 
BEC soliton without dissipation in laboratory.


\end{abstract}

%\date{\today}
 
\maketitle

%\maketitle
%\markboth{thedoi}{thedoi}
%\tableofcontents
%\sloppy 

%\section{Introduction}

Solitons are solutions of wave equation where localization is obtained due
to a nonlinear interaction. Solitons have been noted in optics  
\cite{1},
high-energy physics and water waves  \cite{1}, and also  in
Bose-Einstein
condensates (BEC)  \cite{2x,2y,2z,3,4}. The bright solitons of BEC 
represent
local maxima \cite{3,4,4a0x,4a0y,4a0z},
whereas dark solitons represent local minima  \cite{2x,2y,2z}. In addition 
to
the
observation of an isolated bright soliton in an expulsive potential
\cite{4}, a
number of bright solitons forming a  soliton train was  observed by
Strecker {\it et al.} \cite{3},  where they suddenly turned a repulsive
BEC of
$^7$Li atoms attractive by manipulating an external magnetic field 
near a Feshbach resonance
\cite{4a}. Consequently, the BEC
collapsed, exploded and generated a soliton train which was studied in
detail. Also, a bright vector soliton in a
 repulsive BEC supported by interspecies attraction has  
been studied \cite{interx,intery}.

A soliton or solitary wave by definition propagates over large time
intervals without visible modification of shape which makes it of special
interest. However, a soliton of BEC, or a BEC in general, suffer loss of
atoms due to three-body recombination leading to formation of molecules
\cite{estx,esty,estz}. This means that a BEC soliton will decay and
eventually disappear as it propagates. It would be of interest if an
artificial situation could be created in laboratory with a supply of atoms
so as to compensate for the three-body recombination loss of a BEC soliton
to generate a dynamically-stabilized soliton. To the best of our knowledge
we demonstrate for the first time,  using the
mean-field Gross-Pitaevskii (GP) equation \cite{11}, that such a
dynamically stabilized soliton could indeed be prepared in a radially 
trapped and axially free BEC.  As a strict
soliton appears only in one dimension, we shall be concerned in this paper
with a quasi-one-dimensional BEC soliton in a cigar-shaped trap in a
axially symmetric configuration. 

To demonstrate the presence of a
dynamically-stabilized dissipation-managed BEC soliton, we employ both
time-dependent and time-independent 
analytic perturbation techniques and a complete numerical solution of the
GP equation. The numerical result is found to be in excellent agreement
with that obtained from the perturbation techniques.

Bright solitons are really eigenfunctions of the one-dimensional 
nonlinear
Schr\"odinger (NLS) equation. However, the experimental realization of
bright solitons in trapped attractive cigar-shaped BECs has been possible
under strong transverse binding which, in the case of weak or no axial
binding, simulates the ideal one-dimensional situation for the formation
of bright solitons.  The dimensionless NLS equation in the attractive or
self-focusing case \cite{1}
\begin{equation}\label{nls} 
i \frac{\partial
\phi}{\partial t}+\frac{1}{2}\frac {\partial^2 \phi}{\partial y^2}+
|\phi|^2\phi =0
\end{equation} 
sustains the following bright
 soliton \cite{1}:
\begin{eqnarray}
\phi(y,t)&=&  a \hskip 3pt \mbox{sech}
[a(y-v t)] \times \nonumber \end{eqnarray} 
\begin{eqnarray}\label{DS}   
\times \exp[ivy -i(v^2-a^2)t/2+i\sigma], 
\end{eqnarray}
with three parameters. 

 The parameter $a$ represents the amplitude as well as pulse width, $v$
represents velocity, the parameter $\sigma$ is a phase constant. The
bright soliton profile is easily recognized for $v=0$ as $|\phi
(y,t)|=a\hskip 3pt \mbox{sech} (ay)$.  There have been experimental
\cite{3,4} and theoretical \cite{4a0x,4a0y,4a0z} studies of the formation 
of bright
solitons in a BEC. In view of this, here we study for the first time the
possibility of a dynamically-stabilized dissipation-managed bright BEC
soliton in a quasi-one-dimensional configuration.

In recent times there have been routine experimental studies on the
formation of BEC in the presence of a periodic axial optical-lattice
potential \cite{optlatx,optlaty,optlatz} formed by a polarized 
standing-wave laser
beam. This leads to a different condition of
trapping from the harmonic trap and generates a BEC of distinct
modulation. Hence we
also consider in this paper the modulations of a  dissipation-managed
bright BEC soliton
 in the presence of an optical-lattice potential.

The time-dependent Bose-Einstein condensate wave
function $\Psi({\bf r},\tau)$ at position ${\bf r}$ and time $\tau $
may
be described by the following  mean-field nonlinear GP  
equation
\cite{11,chaos}
\begin{eqnarray}
\biggr[- i\hbar\frac{\partial
}{\partial \tau}
&-&\frac{\hbar^2\nabla_{\bf r}^2   }{2m}
+ V({\bf r})
- g n+i \frac{\Gamma}{2}-\nonumber 
\end{eqnarray}
\begin{eqnarray}\label{a} 
-i\frac{\hbar}{2}K_3 n^2
 \biggr]\Psi({\bf r},\tau)=0, 
\end{eqnarray}
with normalization 
$\int d{\bf r} |\Psi({\bf r},\tau)|^2 = N. $
Here $m$
is
the mass and  $N$ the number of bosonic atoms in the
condensate, $n\equiv  |\Psi({\bf r},\tau)|^2$ is the boson 
probability density,
 $g=4\pi \hbar^2 a/m $ the strength of
inter-atomic attraction, with
$-a$ the atomic scattering length. 
The trap potential with axial symmetry may be written as  $
V({\bf
r}) =\frac{1}{2}m \omega ^2 (\rho^2+\nu^2 z^2)$ where
 $\omega$ and $\nu \omega$ are the angular frequencies in the radial
($\rho$) and axial ($z$) directions with $\nu$ the anisotropy parameter.
The term $K_3$ denotes the three-body recombination loss-rate coefficient
and
$\Gamma$ denotes the constant alimentation of atoms from an external 
source.

For the study of bright  solitons
we shall reduce Eq. (\ref{a})  to a minimal 
 one-dimensional form under the action of stronger radial trapping. 
The one-dimensional form is appropriate for     studying bright 
solitons in the so-called cigar-shaped quasi-one-dimensional geometry
where 
$\nu << 1$. For radially-bound and axially-free solitons we eventually set 
$\nu =0$.

For  $\nu =0$, Eq. (\ref{a})
can be reduced to an effective 
one-dimensional form by considering 
solutions of the type 
$\Psi({\bf r},\tau)=  \Phi(z,\tau)\psi^{(0)}( \rho)$ 
where
\begin{eqnarray}\label{wfx}
|\psi^{(0)}(\rho)|^2&\equiv&
{\frac{m\omega}{\pi\hbar}}\exp\left(-\frac{m
\omega
\rho^2}{\hbar}\right).
\end{eqnarray}
The expression (\ref{wfx})
corresponds to the 
ground state wave function in the radial variable
in the absence of nonlinear 
interactions and
satisfies
\begin{eqnarray}
-\frac{\hbar^2}{2m}\nabla_\rho ^2\psi^{(0)}
+
\frac{1}{2}m\omega^2\rho^2
\psi^{(0)}&=&\hbar\omega
\psi^{(0)},
\end{eqnarray}
with normalization 
\begin{eqnarray}  
2\pi \int_{0}^\infty |\psi^{(0)}(\rho)|^2 \rho d\rho=1.\nonumber 
\end{eqnarray}  
Now the dynamics is carried by $ \Phi(z,\tau)$ and the radial dependence
is
frozen in the ground state $\psi^{(0)}(\rho)$.

Averaging over the radial mode $\psi^{(0)}(\rho)$, 
i.e., multiplying
Eq. (\ref{a}) 
by  $\psi^{(0)*}(\rho)$
and integrating over $\rho$, we obtain the following one-dimensional 
dynamical equation \cite{abdul}:
\begin{eqnarray}\label{i} \biggr[ - i\hbar\frac{\partial
}{\partial t}
-\frac{\hbar^2}{2m}\frac{\partial^2}{\partial z^2}
-  g \frac{m\omega}{2\pi \hbar}
|\Phi|^2
 \nonumber   \end{eqnarray} 
\begin{eqnarray}
+ i \frac{\Gamma}{2}
- i \frac{\hbar}{2}\frac{m^2\omega^2}{3\pi^2\hbar^2}| \Phi|^4
 \biggr] \Phi(z,\tau)=0, 
\end{eqnarray}
In Eq. (\ref{i}) 
the normalization 
is given by 
\begin{eqnarray}
\int_{-\infty}^\infty |\Phi(z,\tau)|^2
dz = N   \nonumber
\end{eqnarray}  
 and  we have set the anisotropy parameter 
$\nu=0$
 to remove the axial trap and thus to generate an axially-free
quasi-one-dimensional soliton.

For calculational purpose it is convenient to reduce 
the set  (\ref{i})  to
dimensionless form 
by introducing convenient  dimensionless variables. 
For this purpose 
we consider the dimensionless variables 
$t=\tau \omega$,
$y=z /l$,
${\phi}=
\sqrt{(2a)} \Phi$, $\gamma=\Gamma/(\hbar \omega)
$, and $\xi= K_3/(24\pi^2a^2l^4 \omega)$
with $l=\sqrt{\hbar/( \omega m)}$, 
so that 
\begin{eqnarray}\label{m} \biggr[  i
\frac{\partial
}{\partial t}
+\frac{1}{2}\frac{\partial^2}{\partial y^2} 
+    
\left|{{\phi}}\right|^2  
 \biggr]{\phi}({y},t)=i \epsilon(\phi)\phi(y,t),
\end{eqnarray}
where 
\begin{equation}\label{ep}
\epsilon(\phi)= \frac{\gamma}{2}
- \xi
  \left|{{\phi}}\right|^4.                  
\end{equation}
and the normalization   is given by 
\begin{eqnarray}
\int_{-\infty}^\infty |\phi(y,t)|^2 dy =\frac{2aN}{l} . \nonumber 
\end{eqnarray}    

In addition to studying an axially-free dissipation-managed soliton, we
also consider a dissipation-managed soliton in a periodic optical-lattice
potential $V(y)= V_0\sin^2(2\pi y/\lambda)$ in the axial 
direction \cite{optlatx,optlaty,optlatz}, so
that the NLS
equation of interest becomes 
\begin{eqnarray}\label{n} \biggr[  i
\frac{\partial
}{\partial \tau}
+\frac{1}{2}\frac{\partial^2}{\partial y^2} 
+    
\left|{{\phi}}\right|^2  -V(y)
 \biggr]{\phi}({y},t)=i \epsilon(\phi)\phi(y,t),
\end{eqnarray}
where $V_0$ is the strength and $\lambda$ is the wavelength of the laser
used to generate the optical-lattice potential.
Equation (\ref{n}) will generate a dissipation-managed soliton with an
axial periodic modulation.

In Eq. (\ref{m}) $\epsilon(\phi)$ is a small perturbation. In the absence
of a perturbation $(\epsilon=0)$ the soliton profile is known. The fate
of the  
soliton when  $\epsilon \ne 0$ can be
obtained by a time-dependent perturbation technique \cite{1} which 
assumes that the functional form of the soliton remains
unchanged in the presence of a small perturbation. However,  the soliton
parameters change with propagation. The most general form of the perturbed
soliton is taken as \cite{1}

\begin{eqnarray} \label{per}
\phi(y,t)= a(t)\mbox{sech}
[a(t)(y-q(t))]\exp[i\sigma(t)-iy\delta(t)].
\end{eqnarray} 
In the absence of a perturbation, $a$ and $\delta$ are
constants but $q(t)$ and $\sigma(t)$ are determined by the following
ordinary differential equations \cite{1}:
\begin{eqnarray}          
\frac{dq}{dt}=-\delta, \quad \quad \frac{d\sigma}{dt}=\frac{1}{2}
(a^2 -\delta^2).
    \end{eqnarray}              
 The use of the  perturbation technique leads to the following 
ordinary differential equations for the soliton parameters $a$ and
$\delta$ \cite{1}:
\begin{eqnarray}\label{pq}         
\frac{da}{dt}&=& \mbox{Re} \int_{-\infty}^\infty \epsilon(\phi)
\phi^*(y)dy, \\
\frac{d\delta}{dt}&=& -\mbox{Im} \int_{-\infty}^\infty \epsilon(\phi)     
\tanh [a(y-q)] \phi^*(y)dy.     
 \end{eqnarray}         
In this work we shall be concerned only with the  profile of the
soliton with zero velocity: $q(t)=0$. This can be obtained by solving
Eq. (\ref{pq}). With the $\epsilon$ of Eq. (\ref{ep}) and the $\phi$ of
Eq. (\ref{per}), Eq. (\ref{pq}) becomes
\begin{eqnarray}
\frac{da}{dt}&=& \int_{-\infty}^\infty\frac{\gamma}{2} |\phi|^2 dy
-\int_{-\infty}^\infty \xi  |\phi|^6 dy     \\
&=& \gamma a - \frac{16}{15}\xi a^5. \label{fi}
 \end{eqnarray}      

We are interested in finding the constant amplitude $a$ of a 
dynamically-stabilized dissipation-managed 
soliton  
for large 
$t$ from the solution of Eq. (\ref{fi}). It is assumed that the
perturbation is switched on at $t =0$. The initial soliton 
at $t =0$ is taken to satisfy 
 $a(t = 0) =a_0$. With this condition
Eq. (\ref{fi}) can be integrated to yield
\begin{eqnarray}        
\ln \biggr[a\biggr(\frac{16}{15}\xi a^4-
\gamma\biggr)^{-1/4}\biggr]_{a=a_0}^a = \gamma t,
\end{eqnarray}          
the solution of which is 
\begin{eqnarray}  
a^4= \frac{15 a_0^4\gamma e^{4\gamma t}}{16 a_0^4\xi e^{4\gamma
t}-(16
\xi a_0^4-15 \gamma)},
    \end{eqnarray}         
which for  finite $\xi$ and $\gamma$ leads for large time $t$ 
to  the amplitude
\begin{eqnarray}   \label{gov}
a= \biggr[ \frac{15 \gamma}{16 \xi} \biggr]^{1/4}.
   \end{eqnarray}       
 The most interesting feature of this result is that the
lowest-order 
perturbation-theory prediction for the amplitude of the
dissipation-managed soliton is independent of the initial choice for the  amplitude
$a_0$ at $t=0$. The dissipation-managed soliton is  robust and depends
only on the ratio $\gamma/\xi$ of the 
dissipation parameters $\gamma$ and $\xi$ and is
independent
of the  initial choice $a_0$ in the lowest-order  perturbation 
theory.

The above result that a robust time-independent soliton of amplitude 
(\ref{gov})
can
be formed at large times in the presence of a weak dissipation  suggests
that the same
 could be derivable from a  time-independent perturbation analysis.
For a ``stationary state" $\phi(y,t)= \phi(y) \exp(i\mu t)$,   from
Eq. (\ref{m}) one could write the
following
time-independent equation
\begin{eqnarray}\label{mm} \biggr[  -\mu
+\frac{1}{2}\frac{\partial^2}{\partial y^2} 
+    
\left|{{\phi}}\right|^2  
 \biggr]{\phi}({y})=i \epsilon(\phi)\phi(y),
\end{eqnarray}
where $\mu $ is a  chemical potential. Employing the usual
time-independent
perturbation theory, for Eq. (\ref{mm}) to have the soliton
(\ref{per}) 
with
$q(t)=0$ as a solution, one should have 
\begin{eqnarray}\label{ti}
\langle \phi  | \epsilon (\phi)| \phi \rangle \equiv
\int_{-\infty}^{\infty}  \epsilon (\phi)|\phi(y)|^2 dy= 0. 
\end{eqnarray}       
With $\epsilon(\phi)$ given by Eq. (\ref{ep}) and $\phi(y)=a\quad
\mbox{sech}(ay)$, the solution of  
Eq. (\ref{ti}) is given by Eq. (\ref{gov}). This demonstrates 
the equivalence between the results of time-dependent and time-independent
perturbation theories. However, it should be recalled that perturbative  
result
(\ref{gov})
is valid only for small dissipation $\epsilon(\phi)$ or for
small 
values of $\gamma$ and $\xi$.

We solve the NLS equation (\ref{m}) for dynamically stabilized
dissipation-managed bright solitons numerically using a time-iteration
method based on the Crank-Nicholson discretization scheme elaborated in
Ref. \cite{sk1x,sk1y}.  We discretize the mean-field-hydrodynamic equation 
using
time step $0.01$ and space step $0.1$.

We performed the time evolution of Eq.  (\ref{m}) starting with the
solution $\phi(y) = 0.5 \mbox{sech}(y/2) \exp(it/8)$ with $a_0=1/2$ for
$\epsilon(\phi) = 0$ at $t=0$. The dissipation $\epsilon(\phi) $ is
introduced for $t > 0$ and we look for the dissipation-managed soliton
for large time $t$.  If a converged solution is obtained, it
corresponds to the dissipation-managed soliton. We also repeated the
calculation with a different initial choice $\phi(y) = \mbox{sech}(y)
\exp(it/2)$ with $a_0=1$ to verify if the dissipation-managed soliton
is independent of the initial choice as predicted by the perturbation
treatment above.

\begin{figure}%[!ht]
 
\begin{center}
\includegraphics[width=1.\linewidth]{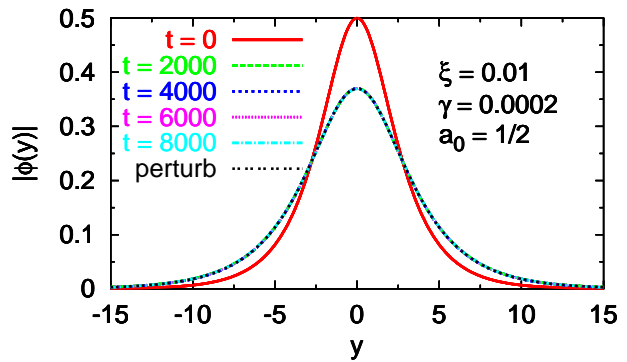}
\end{center}

\caption{(Color online)  The dissipation-managed
dynamically-stabilized soliton $|\phi(y)|$
of the NLS equation (\ref{m}) vs $y$ at different times for 
$\xi = 0.01, \gamma=0.0002$ and $a_0 = 1/2$. 
The initial soliton at $t=0$ is taken as $\phi(y)= 0.5 \mbox{sech}(x/2)$. 
The displayed perturbed solution labelled `perturb'
is in excellent agreement with the numerical result.
}
\end{figure}

\begin{figure}%[!ht]
 
\begin{center}
\includegraphics[width=1.\linewidth]{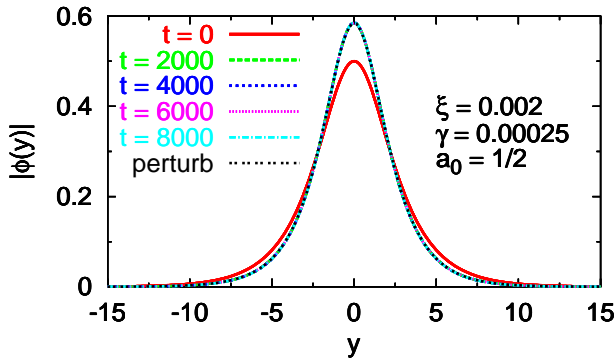}
\end{center}

\caption{(Color online)  
Same as in Fig. 1 but with $\xi =0.002$ and $\gamma= 0.00025$. 
}

\end{figure}

In the first numerical simulation we take $\xi =0.01, 
\gamma =0.0002$ and $a_0=1/2$. The $t=0$ solution is taken as 
$\phi(y) = 0.5\mbox{sech}(y/2)$. Upon a time evolution of the NLS
equation (\ref{m}) the dissipation-managed soliton emerges at large times. 
In Fig. 1 we plot the numerically-calculated 
dissipation-managed soliton  at times
$t=2000, 4000, 6000,$ and 8000 as
well as the result of the lowest-order perturbation theory.
At large times the  numerically-calculated          
dissipation-managed soliton exhibits excellent
convergence properties and agrees very well with the  result of the 
lowest-order perturbation theory. We also repeated this calculation with a
different  $a_0 (=1)$. In agreement with the result of perturbation
theory,
the dissipation-managed soliton for $a_0 =1$ is indistinguishable from
that for $a_0=1/2$. Although the  dissipation-managed soliton appears like
a stable solution of the NLS equation (\ref{m}), it cannot be called a
stationary solution of a time-independent Schr\"odinger-type equation with
real eigenvalue. It is rather a dynamically-stabilized solution of the
dissipative NLS equation  (\ref{m}) where there is a delicate cancellation
between the two imaginary dissipative terms $\xi$ and $\gamma$ to
maintain a stable profile of the soliton.

To study how the dissipation-managed soliton changes with a change of the
dissipation parameters $\xi $ and $\gamma$ we repeat the calculation of
Fig. 1 first with smaller values of these parameters:  $\xi = 0.002$ and
$\gamma = 0.00025$. A very similar scenario emerges for the
dynamically-stabilized soliton which we exhibit in Fig. 2. However, in
this case the dynamically-stabilized soliton has a 
larger amplitude $a (=0.58) $
governed by Eq. (\ref{gov}) 
than in Fig. 1 where   $a (=0.37) $.  In these numerical studies we
verified that for small
dissipation the  dissipation-managed soliton is
independent of $a_0$  and is solely determined by the ratio $\gamma/\xi$. 
However, for large dissipation     parameters this is not quite so and
quite
expectedly the  numerical solution of the soliton does not agree with
the result of lowest-order perturbation theory. Also, with the increase of 
dissipation parameters it  becomes increasingly difficult numerically to
find a stabilized soliton. With further increase in dissipation parameters
we could not obtain a stabilized soliton. Numerically, no stabilized
soliton could be found for $\xi =0.1$.

%In our numerical investigation as in the theoretical study of Refs.
%\cite{jz,ska} we use $\omega = 2\pi \times 100$ Hz, and take $m_B$ as the
%mass of $^{87}$Rb. Consequently, the unit of length $l\approx1$ $\mu$m
%and unit of time $2/\omega \approx 3$ ms.

\begin{figure}%[!ht]
 
\begin{center}
\includegraphics[width=1.\linewidth]{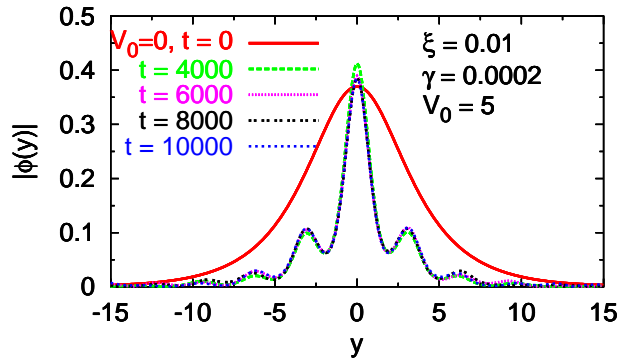}
\end{center}

\caption{(Color online)  The dissipation-managed soliton of Eq. (\ref{n})
in the presence of an optical-lattice potential with $V_0=5$ and 
$\lambda = 2\pi$. The other parameters are the same as in Fig. 1.  
} \end{figure}

Finally, we consider a dissipation-managed  bright soliton formed on a
periodic
optical-lattice potential controlled by Eq. 
(\ref{n}).
In our simulation we take $V_0=5$ and $\lambda = 2\pi$. To
solve Eq.  (\ref{n}) with this optical-lattice potential 
we take   the
initial soliton at $t= 0$ to be  the one calculated in Fig. 1 
above for $V(y)=0$ and
consider the time
evolution of Eq. (\ref{n}). During this time evolution  the
optical-lattice potential is slowly introduced in an interval of time $t$
of about 2000 units,  so that a stable dissipation-managed 
soliton  of Eq. (\ref{n}) is obtained at large
time. We study the stability of this dissipation-managed soliton by
continuing the time evolution. Again a dynamically-stabilized
dissipation-managed soliton is obtained which is independent of the input
guess for $a_0$. 

Now  let us see to what values of the three-body recombination rate
$K_3$ the parameters $\xi =0.01 $ and 0.002 considered in this
paper correspond to for  typical experimental values of $a = 10$ nm, $l = 
1$ $\mu$m  and
$\omega = 2\pi\times 50$ Hz. Recalling that $K_3= 24 \pi^2 \xi  a^2
l^4\omega$ we find that for $\xi =0.01$, $K_3=7.5 \times 10^{-26}$
cm$^6$/s
and for $\xi =0.002$, $K_3=1.5 \times 10^{-26}$ cm$^6$/s. Typical
estimates of $K_3 $ for different atoms \cite{estx,esty,estz} are 
comparable to or
less
than these
values.  Hence the present demonstration that dissipation-managed soliton
can be obtained for typical values of $\xi$ less than 0.01 is compatible
with  experimental values of $K_3$. This leads to the conclusion that
dissipation-managed solitons can be generated in laboratory  for
realistic values of the parameters.

To summarize, we have demonstrated the possibility of the generation of
a dynamically-stabilized dissipation-managed robust soliton in a
quasi-one-dimensional BEC under a typical experimental situation. This can
be used in laboratory to generate robust BEC solitons where the loss of
atoms due to three-body recombination could be compensated by alimentation
of atoms from an external source.

\acknowledgments
%\begin{acknowledgement}

{The work is supported in part by the CNPq and FAPESP
of Brazil.}
 
%\end{acknowledgement}
  
%\section*{References}

\end{document}